\documentclass[lettersize,journal]{IEEEtran}
\usepackage{amsmath,amsfonts}
\usepackage{algorithmic}
\usepackage{algorithm}
\usepackage{circuitikz}
\usepackage{amssymb}
\usepackage{array}
\usepackage{booktabs}
\usepackage{textcomp}
\usepackage{multirow}
\usepackage{stfloats}
\usepackage{url}
\usepackage{circuitikz}
\usetikzlibrary{arrows.meta,positioning,fit}
\usetikzlibrary{arrows.meta,positioning,calc}
\tikzset{>=Stealth}
\usepackage{verbatim}
\usepackage{graphicx}
\usepackage{cite}
\usepackage{orcidlink}
\hypersetup{hidelinks} 
\usepackage{hyperref}
\usepackage{subcaption}  
\usepackage{booktabs} 

\begin{document}
\title{A Physics-Based Digital Human Twin for Galvanic-Coupling Wearable Communication Links}

\author{Silvia~Mura\,\orcidlink{0000-0002-0207-5730},
Chiara~Cavigliano\,\orcidlink{0009-0002-6455-6676},~Anna~Marcucci\,\orcidlink{0009-0009-2755-8489},~Pietro~Savazzi\,\orcidlink{0000-0003-0692-8566},~Anna~Vizziello\,\orcidlink{0000-0002-6378-141X},~Maurizio~Magarini\,\orcidlink{0000-0001-9288-0452}%
}

\maketitle
\begin{abstract}
This paper presents a systematic characterization of wearable galvanic coupling (GC) channels under narrowband and wideband operation. A physics-consistent digital human twin maps anatomical properties, propagation geometry, and electrode–skin interfaces into complex transfer functions directly usable for communication analysis. Attenuation, phase delay, and group delay are evaluated for longitudinal and radial configurations, and dispersion-induced variability is quantified through attenuation ripple and delay standard-deviation metrics versus bandwidth. Results confirm electro–quasistatic, weakly dispersive behavior over 10~kHz--1~MHz. Attenuation is primarily geometry-driven, whereas amplitude ripple and delay variability increase with bandwidth, tightening equalization and synchronization constraints. Interface conditioning (gel+foam) significantly improves amplitude and phase stability, while propagation geometry governs link budget and baseline delay. Overall, the framework quantitatively links tissue electromagnetics to waveform distortion, enabling informed trade-offs among bandwidth, interface design, and transceiver complexity in wearable GC systems.
\end{abstract}


\maketitle
\section{Introduction}\label{sec:intro}

Wearable medical systems demand ultra-reliable, energy-efficient, and secure communication to support continuous monitoring, real-time processing, and closed-loop actuation. Unlike conventional wireless networks, they operate under tight energy budgets, strict safety limits on injected currents, and highly variable propagation conditions shaped by tissue composition, hydration, motion, and inter-subject variability. The physical layer must therefore be designed to ensure robustness, safety compliance, and stable performance in dynamic biological environments.

Intra-body Networks (IBNs) have emerged as a key connectivity paradigm for the Internet of Medical Things (IoMT) and the Internet of Bio-Nano Things (IoBNT), enabling communication among distributed on-body and implantable devices~\cite{naranjo2018past},\cite{akyildiz2014terahertz}. Intra-body Communication (IBC) exploits biological tissues as the transmission medium and can be categorized into mechanical, electromagnetic, and electrical-coupling approaches~\cite{yaghoubi2022wireless}. Mechanical techniques (e.g., ultrasound) support deep-tissue propagation but are sensitive to multipath and tissue inhomogeneity, whereas electromagnetic solutions, especially at millimeter-wave and terahertz frequencies, provide high data rates at the cost of severe absorption in water-rich tissues~\cite{akyildiz2014terahertz}.

Field-based coupling mechanisms offer a more energy-efficient alternative with inherently reduced radiation leakage, including inductive (magnetic), capacitive (electric-field), and galvanic (conductive electric-field) coupling. Inductive coupling relies on near-field magnetic flux linkage and is sensitive to coil alignment and separation, limiting robustness in dynamic wearable scenarios. Capacitive coupling exploits electric-field interaction but typically requires a stable external ground reference, making it vulnerable in floating wearable configurations. In contrast, galvanic coupling (GC) injects differential currents directly into conductive tissues, creating a confined conduction path within the body. This enables localized propagation, improved power transfer efficiency, and minimal radiation losses~\cite{wegmueller2009galvanic,tomlinson2018comprehensive}, making GC well suited for short-range, energy-constrained wearable links.

Galvanic coupling systems typically operate in the 10~kHz--1~MHz range, where channel behavior is governed by tissue conductivity, electrode geometry and spacing, and electrode--skin interface impedance~\cite{wegmueller2010signal}. Although attenuation and transfer functions have been reported~\cite{wegmueller2009galvanic}, the literature remains fragmented from a communication-system perspective. Most studies focus on magnitude response, overlooking phase and group delay, which critically affect dispersion and synchronization in broadband links. In addition, simulations and measurements are rarely rigorously calibrated, and narrowband circuit models are seldom reconciled with wideband field descriptions, limiting reproducibility and hindering unified system-level design guidelines.

To address these limitations, this work introduces a physics-based digital human twin for galvanic wearable communication that integrates finite-element modeling with measurement-driven calibration within a coherent and reproducible framework. By consistently linking electromagnetic field behavior to communication-level channel metrics, the proposed approach enables physically grounded channel characterization and supports the systematic design and optimization of next-generation wearable networks.
\subsection{Related Work}

Galvanic coupling has been studied for over a decade as a low-power solution for short-range wearable and implant-adjacent communication, with surveys highlighting the need for physically grounded, reproducible, and safety-aware channel models for medical-grade systems~\cite{naranjo2018past}. Unlike conventional conductors, the human body is a heterogeneous, dispersive volume whose electrical properties vary with frequency, anatomy, and physiological state, making accurate yet design-oriented channel modeling inherently challenging.

From an electromagnetic standpoint, biological tissues are commonly described through frequency-dependent dielectric models, most notably the Cole–Cole formulation~\cite{gabriel1996dielectric}. This representation has enabled extensive finite-element (FE) investigations of GC propagation~\cite{xu2012fe_wholebody_gcibc,ahmed2019simulation,callejon2014finiteelement,modak2022biophysical}, clarifying the influence of fat thickness, muscle conductivity, and electrode placement on attenuation and current distribution. Nevertheless, the majority of simulation-based studies concentrate on path loss or equivalent impedance, while phase response and dispersion-related metrics, such as phase and group delay, are rarely reported in a systematic manner. Consequently, the impact of tissue dispersion on waveform distortion and broadband link integrity remains only partially quantified.

Complementary analytical and distributed-parameter models have provided valuable physical insight into current return paths and quasi-static propagation mechanisms~\cite{multipath2015_gcibc,callejon2012distributed,pun2011quasistatic}. While computationally efficient, these formulations typically rely on simplified geometries and idealized boundary conditions. Electrode structure, material properties, and electrode–skin interface dynamics are often captured through lumped approximations rather than explicitly modeled configurations, limiting their extension toward configuration-aware optimization or subject-specific validation.

Experimental investigations have further demonstrated that electrode placement, spacing, and interface conditioning critically affect GC channel stability and repeatability~\cite{wegmueller2009galvanic,callejon2014finiteelement}. Ag/AgCl electrodes with conductive gel improve measurement consistency by reducing interface impedance variability~\cite{wegmueller2010signal,callejon2013galvanic}, and recent wearable electrode designs emphasize hydrogel composition, mechanical conformity, and hydration control for long-term robustness~\cite{martinez2023hydrogel,takagi2024wearable,fu2020dry}. Despite these advances, simulations and measurements are frequently treated as separate validation steps. Quantitative alignment is often limited to qualitative trend comparison, and structured parameter-level calibration remains uncommon. This separation restricts the development of predictive channel representations suitable for systematic design-space exploration.

At the system level, circuit-aware analyses have shown that front-end impedance and receiver topology reshape the observed transfer function~\cite{chen2020circuit,maity2019biophysical}, while energy-efficient IBC surveys advocate cross-layer evaluation frameworks~\cite{survey2021_energy_ibc}. Yet most prior contributions still focus primarily on attenuation or impedance magnitude, with limited characterization of the full complex transfer function. Without structured treatment of phase and group delay, it is difficult to translate electromagnetic modeling into waveform-level guidelines or to quantify dispersion-induced distortion in wideband operation.

In parallel, the digital-twin (DT) paradigm has emerged in biomedical cyber-physical systems to couple physics-based models with experimental data for predictive optimization~\cite{humanbodydt2023_masterplan}. While FE simulations and measurements are common in GC research, they are seldom integrated into a unified, calibration-driven surrogate model statistically aligned with experiments and directly applicable to communication-system design. Such design-oriented integration is established in alternative-physics communication domains, including molecular and multi-scale systems~\cite{guo2015molecular,baydas2023mimo}, but remains largely underdeveloped for GC-based wearable links.
Collectively, prior research has established strong foundations in tissue electromagnetics, electrode-interface behavior, and experimental channel characterization. What remains missing is an integrated framework that simultaneously (i) captures dispersion-aware channel responses including phase and group delay, (ii) explicitly models realistic electrode and interface configurations, (iii) aligns simulations with measurements through structured parameter calibration, and (iv) exposes the resulting channel representation in a form directly usable for waveform and system-level optimization. Addressing this gap requires moving beyond isolated modeling or measurement studies toward a unified, validated digital representation of the GC channel capable of guiding next-generation wearable communication design.
\subsection{Contributions}
Our prior work in~\cite{CaviglianoGCWearable} characterized attenuation in wearable GC links, demonstrating feasibility and magnitude-based trends, but did not address radial propagation, dispersion, bandwidth-dependent distortion, or system-level implications.

This study extends that foundation through a physics-consistent digital human twin combining electro–quasistatic modeling with communication-theoretic channel metrics. By reconstructing the full complex transfer function, the framework enables joint analysis of attenuation, phase delay, and group delay, quantitatively linking tissue electromagnetics, interface conditions, geometry, and bandwidth to waveform integrity and transceiver design.

The main contributions are:

\begin{itemize}
\item {Physics-based digital human twin:} A calibrated finite-element framework producing complex GC transfer functions and dispersion-aware communication metrics.
\item {Narrowband and wideband integration:} Unified analysis of attenuation and delay behavior across carrier frequency and signal bandwidth.
\item {Dispersion quantification:} Systematic evaluation of attenuation ripple and delay variability as functions of bandwidth.
\item {Interface impact isolation:} Experimental validation under clinically relevant Ag/AgCl gel+foam conditions.
\item {Design guidelines:} Explicit recommendations for bandwidth selection, carrier frequency, interface conditioning, and equalization complexity in wearable GC systems.
\end{itemize}

\subsection{Organization}

The remainder of the paper is organized as follows. Section~\ref{sec:model} introduces the channel model, Sect.~\ref{sec:tissue_properties} and ~\ref{sec:electrodes}  presents tissue and interface characterization, Sect.~\ref{sec:digital_twin} describes the digital human twin and validation, Sect.~\ref{sec:num_results} describes the numerical results, and Sect.~\ref{sec:conclusion} concludes the paper.
\section{System Model}\label{sec:model}

\begin{figure}[!t]
    \centering
    \includegraphics[width=\columnwidth]{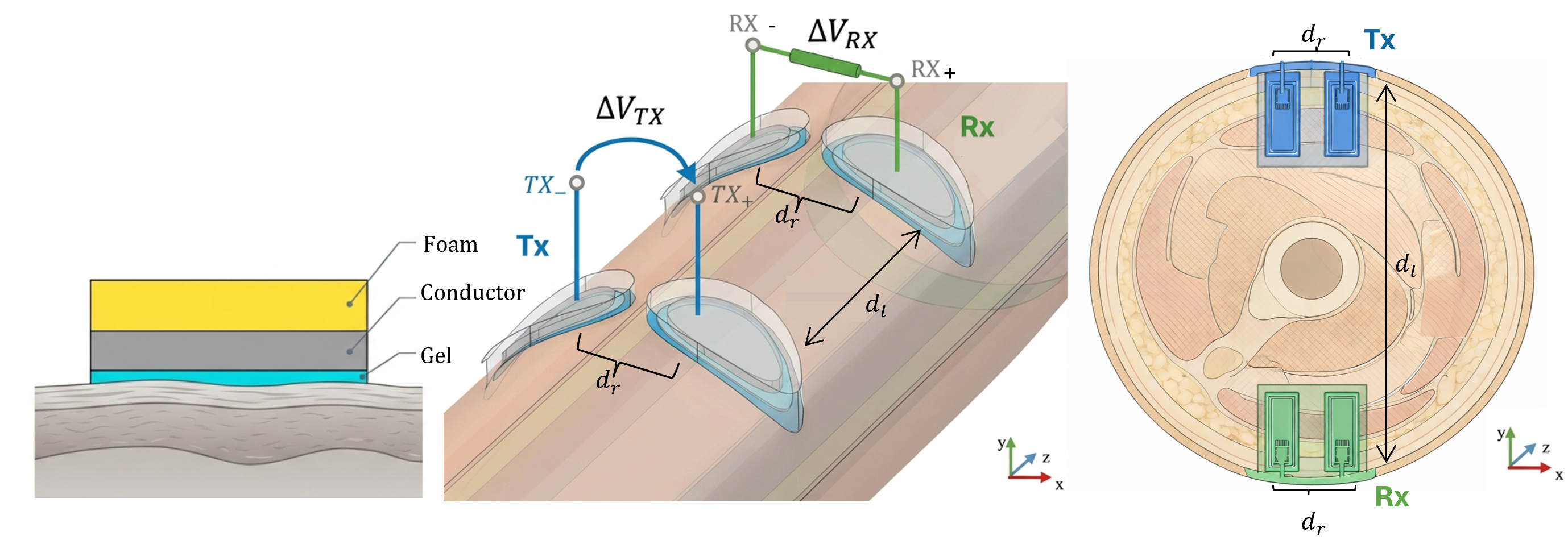}
    \caption{Galvanic-coupling wearable link model showing the layered electrode structure (left), longitudinal Tx–Rx configuration with spacing $d_r$ and separation $d_l$ (center), and radial arm cross-section with tissue layers (right).}
    \label{fig:systemModel}
\end{figure}

We consider a GC wearable link realized by two differential electrode pairs placed on the arm surface, namely a transmitter (Tx) electrode and a receiver (Rx) electrode (Fig.~\ref{fig:systemModel}). The intra-pair spacing within each electrode is denoted by $d_r$, while the inter-pair separation is denoted by $d_l$ when electrodes are aligned along the arm axis. Two geometries, representative of practical wearable deployments, are considered:
\begin{enumerate}
    \item \textit{Longitudinal configuration:} Tx and Rx dipoles are separated by $d_l$ along the arm axis and share comparable circumferential position; the intra-pair spacing is $d_r$.
    \item \textit{Radial configuration:} Tx and Rx dipoles are displaced circumferentially (i.e., across the arm perimeter), representative of armband-like placements; geometry is parameterized by $d_r$ and by the angular separation between Tx and Rx dipoles.
\end{enumerate}
In both cases, the electrical propagation is governed by conductive currents in dispersive biological tissues, and the electrode--skin interface plays a dominant role in determining the effective injection and sensing impedances.

\subsection{Transmit Signal and Current Injection}\label{subsec:tx_wideband}

To connect the physical GC link with a communication-theoretic abstraction, 
the transmitter is modeled as generating a complex baseband waveform that is converted 
into a differential current injected across the Tx electrodes.

Let $x[k]\in\mathbb{C}$ denote the discrete-time symbol sequence 
(data or probing), and let $p(t)$ be a unit-energy pulse-shaping waveform. 
The continuous-time complex envelope is
\begin{equation}
s(t)=\sum_{k} x[k]\,p(t-kT),
\label{eq:st}
\end{equation}
where $T$ is the symbol interval. For channel sounding, $x[k]$ is a known deterministic training sequence, 
whereas for data transmission it represents symbols drawn from a finite complex constellation. The Tx front-end is modeled as a linear time-invariant (LTI) transconductance stage with impulse response $g_I(t)$. 
The injected differential current is therefore
\begin{equation}
i_{\mathrm{TX}}(t) = (g_I * s)(t),
\label{eq:itx_time}
\end{equation}
where $*$ denotes convolution. Since the front-end is LTI, its behavior is equivalently described in the frequency domain
\begin{equation}
I_{\mathrm{TX}}(f) = G_I(f)\, S(f),
\label{eq:itx_freq}
\end{equation}
with $G_I(f) = \mathcal{F}\{g_I(t)\}$ denoting the effective transconductance (A/V). For wearable GC systems operating in 
$\mathcal{B} \approx 10\,\mathrm{kHz}–1\,\mathrm{MHz}$,
the current driver bandwidth is designed significantly larger than $\mathcal{B}$, 
so that $G_I(f)$ can be assumed approximately flat over the operating band\cite{callejon2012distributed}. 
Under this condition, the injected current spectrum is shaped primarily by $S(f)$.

The differential voltage at the injection port is
\begin{equation}
\Delta V_{\mathrm{TX}}(f)
=Z_{\mathrm{TX}}(f)\,I_{\mathrm{TX}}(f),
\label{eq:vtx}
\end{equation}
where $Z_{\mathrm{TX}}(f)$ is the complex input impedance seen at the Tx electrodes. 
Within $\mathcal{B}$, the quasi-static approximation holds~\cite{gabriel1996dielectric,pun2011quasistatic}, 
so electromagnetic wave effects are negligible and tissue transport is governed by conduction, and displacement currents. 
Under these conditions, the medium is linear and time-invariant, which justifies frequency-domain characterization and ensures that time-domain distortion can be obtained by inverse Fourier transformation.

The input impedance is modeled as
\begin{equation}
Z_{\mathrm{TX}}(f)=Z_{e,\mathrm{TX}}(f)+Z_{\mathrm{body,TX}}(f),
\label{eq:Zin}
\end{equation}
where $Z_{e,\mathrm{TX}}(f)$ captures electrode-interface effects and $Z_{\mathrm{body,TX}}(f)$ represents the local tissue contribution.

\subsection{GC Channel, Transfer Impedance, and Equivalent Circuit}\label{subsec:sysmodel_q1}

Within the frequency band of interest $\mathcal{B}$, the GC link is modeled as a LTI two-port system characterized by the measurable differential voltage transfer function
\begin{equation}
\label{eq:channel}
H(f)=\frac{\Delta V_{\mathrm{RX}}(f)}{\Delta V_{\mathrm{TX}}(f)}
=\alpha(f)e^{-j\phi(f)},
\end{equation}
where $\Delta V_{\mathrm{TX}}(f)$ and $\Delta V_{\mathrm{RX}}(f)$ denote the differential Tx and Rx voltages, respectively. The magnitude $\alpha(f)$ describes frequency-dependent attenuation, while $\phi(f)$ represents the corresponding phase rotation introduced by conductive transport through biological tissue. This definition is directly consistent with experimental practice, since both voltages are measurable quantities (or computable via FE modeling), thereby ensuring strict alignment between simulation and measurement. 

To isolate the intrinsic conductive interaction between the Tx and Rx dipoles, the open-circuit differential transimpedance is defined as
\begin{equation}
\label{eq:Ztr_def}
Z_{\mathrm{tr}}(f)\triangleq
\frac{\Delta V_{\mathrm{RX}}(f)}{I_{\mathrm{TX}}(f)}
\Big|_{Z_L\rightarrow\infty},
\end{equation}
where $I_{\mathrm{TX}}(f)$ denotes the injected differential current and $Z_L(f)$ is the receiver load impedance. The limiting condition $Z_L\rightarrow\infty$ enforces open-circuit reception and eliminates receiver loading effects, ensuring that $Z_{\mathrm{tr}}(f)$ represents the intrinsic mutual impedance of the dispersive multilayer volume conductor~\cite{callejon2012distributed}. In this sense, $Z_{\mathrm{tr}}(f)$ depends solely on anatomical geometry, electrode placement, tissue electromagnetic properties, and environmental parasitic couplings, and therefore constitutes a channel parameter independent of the receiver electronics.

For physical interpretability, the transimpedance may be expressed as
\begin{equation}
\label{eq:Ztr_decomp}
Z_{\mathrm{tr}}(f)=
Z_{\mathrm{spread,TX}}(f)+
Z_{\mathrm{bulk}}^{(\Gamma)}(f)+
Z_{\mathrm{spread,RX}}(f)+
Z_{\mathrm{par}}(f),
\end{equation}
where the spreading terms describe local current redistribution beneath the respective dipoles, $Z_{\mathrm{bulk}}^{(\Gamma)}(f)$ captures volume conduction along the dominant current streamline $\Gamma$ connecting the Tx and Rx regions, and $Z_{\mathrm{par}}(f)$ accounts for parasitic and leakage paths such as surface shunting and distributed capacitive coupling. This decomposition separates local interface phenomena from bulk conductive transport while preserving linear superposition.

Within $\mathcal{B}$, the quasi-static approximation is valid~\cite{gabriel1996dielectric,pun2011quasistatic}, such that electromagnetic wave propagation effects are negligible and transport is governed by conductive and displacement currents in a heterogeneous, dispersive medium. The complex impedance of tissue along the streamline $\Gamma$ can therefore be compactly represented as
\begin{equation}
\label{eq:Zbulk}
Z_{\mathrm{bulk}}^{(\Gamma)}(f)=
\int_{\Gamma}
\frac{1}{\sigma(\ell,f)+j2\pi f\,\varepsilon(\ell,f)}
\,\mathrm{d}\ell,
\end{equation}
where $\sigma(\ell,f)$ and $\varepsilon(\ell,f)=\varepsilon_0\varepsilon_r(\ell,f)$ denote the spatially and frequency-dependent conductivity and permittivity, respectively. These parameters vary across anatomical layers such as skin, fat, muscle, and bone, and follow dispersive models including Cole--Cole parameterizations~\cite{gabriel1996dielectric}. Equation~\eqref{eq:Zbulk} makes explicit that dispersion in $\sigma$ and $\varepsilon$ induces both amplitude attenuation and phase rotation, thereby shaping the frequency-dependent group delay of the channel. Electrode placement modifies the effective streamline $\Gamma$, altering the relative contribution of superficial and deep conductive pathways and consequently the overall transfer characteristics.

To incorporate electrode–tissue interface effects, the differential electrode impedance at port $k\in\{\mathrm{TX},\mathrm{RX}\}$ is modeled as
\begin{equation}
\label{eq:Ze_full}
Z_{e,k}(f)=
R_c+
\frac{1}{j2\pi f C_{dl}}+
\frac{t_g}{A\left(\sigma_g+j2\pi f\varepsilon_g\right)}+
\frac{1}{j2\pi f C_{\mathrm{par}}(f)},
\end{equation}
where $R_c$ denotes the effective charge-transfer resistance and $C_{dl}$ the double-layer capacitance. The conductive gel layer of thickness $t_g$ and area $A$ is modeled as a dispersive slab with complex admittance $\sigma_g+j2\pi f\varepsilon_g$, while $C_{\mathrm{par}}(f)$ captures parasitic capacitances associated with foam backing and nearby conductors~\cite{wegmueller2010signal,callejon2013galvanic}. Conductive gel primarily increases mid-band interface admittance and reduces $R_c$, thereby improving injection efficiency. Foam backing stabilizes mechanical pressure and hydration, reducing temporal variability of the contact impedance and slightly modifying parasitic capacitances. These interface effects directly influence the transmitter input impedance $Z_{\mathrm{TX}}(f)$ and indirectly affect $Z_{\mathrm{tr}}(f)$ through modified boundary conditions and altered current distributions within the surrounding tissue.

Defining the differential self-impedance at the Rx as
\begin{equation}
\label{eq:Zrx_def}
Z_{\mathrm{RX}}(f)=
Z_{e,\mathrm{RX}}(f)+Z_{\mathrm{body,RX}}(f),
\end{equation}
where $Z_{\mathrm{body,RX}}(f)$ represents the local tissue contribution at the receiver location, and considering practical operation in which the Rx dipole is terminated by the receiver input impedance $Z_L(f)$, the loaded received voltage is given by
\begin{equation}
\label{eq:Vrx_loaded}
\Delta V_{\mathrm{RX,load}}(f)=
\frac{Z_L(f)}{Z_L(f)+Z_{\mathrm{RX}}(f)}
\,Z_{\mathrm{tr}}(f)\,I_{\mathrm{TX}}(f).
\end{equation}
The prefactor accounts for the voltage division between the intrinsic Rx-port impedance and the receiver load, thereby quantifying loading effects introduced by the measurement instrument or analog front-end.

The measurable equivalent transfer impedance is therefore defined as
\begin{equation}
\label{eq:Zeq_def}
Z_{\mathrm{eq}}(f)\triangleq
\frac{Z_L(f)}{Z_L(f)+Z_{\mathrm{RX}}(f)}\,Z_{\mathrm{tr}}(f),
\end{equation}
so that the observable channel transfer function can be written compactly as
\begin{equation}
\label{eq:H_Zeq}
H(f)=
\frac{\Delta V_{\mathrm{RX,load}}(f)}{\Delta V_{\mathrm{TX}}(f)}
=
\frac{Z_{\mathrm{eq}}(f)}{Z_{\mathrm{TX}}(f)}.
\end{equation}
For high-impedance reception satisfying $|Z_L(f)|\gg|Z_{\mathrm{RX}}(f)|$ over $\mathcal{B}$, the loading factor approaches unity and $Z_{\mathrm{eq}}(f)\approx Z_{\mathrm{tr}}(f)$, such that the measured response closely approximates the intrinsic channel transimpedance.

\begin{figure}[!t]
\centering
\resizebox{\columnwidth}{!}{%
\begin{tikzpicture}[font=\small, line cap=round, line join=round]
\tikzset{
  >=Stealth,
  wire/.style={line width=0.9pt},
  block/.style={draw, rounded corners=2pt, line width=0.9pt, minimum width=26mm, minimum height=10mm, align=center},
  pot/.style={<->, line width=0.9pt},
  itx/.style={->, line width=0.9pt, red!75!black},
}

\coordinate (G)    at (0,0);
\coordinate (TXp)  at (0,3.2);

\coordinate (ZTXL) at (2.2,3.2);
\coordinate (ZTXR) at (4.9,3.2);

\coordinate (VOCL) at (6.2,3.2);
\coordinate (VOCR) at (9.0,3.2);

\coordinate (ZRXL) at (10.2,3.2);
\coordinate (ZRXR) at (12.9,3.2);

\coordinate (RXdown) at (12.9,0.9);

\begin{scope}
  \draw (G) node[ground]{};
  \draw (G) to[isource, l=$I_{\mathrm{TX}}(f)$] (TXp);
\end{scope}

\draw[wire] (TXp) -- (ZTXL);
\node[block] (ZTX) at ($(ZTXL)!0.5!(ZTXR)$) {$Z_{\mathrm{TX}}(f)$};
\draw[wire] (ZTXR) -- (VOCL);

\node[block] (VOC) at ($(VOCL)!0.5!(VOCR)$)
{$V_{\mathrm{oc}}(f)$\\[-1pt]\scriptsize $=Z_{\mathrm{tr}}(f)\,I_{\mathrm{TX}}(f)$};
\draw[wire] (VOCR) -- (ZRXL);

\node[block] (ZRX) at ($(ZRXL)!0.5!(ZRXR)$) {$Z_{\mathrm{RX}}(f)$};

\draw[wire] (ZRXR) -- ++(0.6,0);               
\coordinate (ZLtop) at ($(ZRXR)+(0.6,0)$);
\begin{scope}
  \draw (ZLtop) to[R, l_=$Z_L(f)$] ($(ZLtop |- RXdown)$);
\end{scope}
\coordinate (ZLbot) at ($(ZLtop |- RXdown)$);

\draw[wire] (ZLbot) -- ($(G)+(0,0.9)$) -- (G);

\draw[itx] ($(TXp)+(0.55,0.35)$) -- ++(1.2,0)
  node[midway, above] {\scriptsize $I_{\mathrm{TX}}$};


\draw[pot] ($(ZTXL)+(0,1.05)$) to[out=10,in=170] ($(ZTXR)+(0,1.05)$);
\node at ($(ZTXL)!0.5!(ZTXR)+(0,1.42)$) {$\Delta V_{\mathrm{TX}}(f)$};

\draw[pot] ($(ZRXL)+(0,1.05)$) to[out=10,in=170] ($(ZRXR)+(0,1.05)$);
\node at ($(ZRXL)!0.5!(ZRXR)+(0,1.42)$) {$\Delta V_{\mathrm{RX}}(f)$};

\coordinate (Vtop) at ($(ZLtop)+(1.25,0)$);
\coordinate (Vbot) at ($(ZLbot)+(1.25,0)$);

\draw[pot]
  (Vtop) to[out=-70,in=70] (Vbot);

\node[rotate=90] 
  at ($(Vtop)!0.5!(Vbot)+(0.6,0)$) 
  {$\Delta V_{\mathrm{RX,load}}(f)$};

\node[align=center, font=\footnotesize] at ($(VOC.south)!0.5!(ZRX.south)+(0,-1.35)$)
{$\displaystyle
\Delta V_{\mathrm{RX,load}}(f)=
\frac{Z_L(f)}{Z_L(f)+Z_{\mathrm{RX}}(f)}\,Z_{\mathrm{tr}}(f)\,I_{\mathrm{TX}}(f)$};

\end{tikzpicture}%
}
\caption{Equivalent circuit with aggregated transmitter impedance $Z_{\mathrm{TX}}(f)$, open-circuit voltage $V_{\mathrm{oc}}(f)=Z_{\mathrm{tr}}(f)I_{\mathrm{TX}}(f)$, aggregated receiver impedance $Z_{\mathrm{RX}}(f)$, and load $Z_L(f)$. The red arrow denotes the injected current $I_{\mathrm{TX}}(f)$, while curved arrows indicate the differential voltages.}
\label{fig:equiv_circuit_redone}
\end{figure}
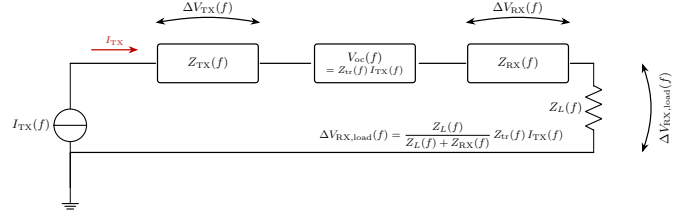

To express attenuation in decibel form while preserving physical interpretability, an impedance-referenced path loss is defined as
\begin{equation}
\label{eq:pathloss}
\mathrm{PL}(f)=
20\log_{10}\!\left(1+\left|\frac{Z_{\mathrm{eq}}(f)}{Z_{\mathrm{ref}}}\right|\right),
\end{equation}
where $Z_{\mathrm{ref}}$ is a chosen reference impedance. A natural measurement-oriented choice is $Z_{\mathrm{ref}}=Z_L(f)$, which directly relates the path loss to the voltage delivered to the receiver input. Importantly, $Z_L(f)$ is a receiver parameter rather than a tissue property; it may be approximately constant for high-impedance instrumentation or frequency-dependent for practical analog front-ends.
Figure \ref{fig:equiv_circuit_redone} details the equivalent circuit.

\subsection{Received Signal}
Under the LTI channel model introduced in the previous subsection, the received differential voltage is expressed as the convolution between the transmitted waveform and the channel impulse response (CIR), corrupted by noise and interference:
\begin{equation}
y(t)=\big(h(t)\ast s(t)\big)+n(t),
\label{eq:rx}
\end{equation}
where $h(t)=\mathcal{F}^{-1}\{H(f)\}$ is the impulse response associated with the transfer function $H(f)$, and $n(t)$ accounts for thermal noise and external interference (e.g., power-line components).

After downconversion and sampling, the equivalent discrete-time baseband model becomes
\begin{equation}
y[k]=\sum_{\ell=0}^{L-1} h[\ell]\,x[k-\ell]+w[k],
\label{eq:discrete}
\end{equation}
where $h[\ell]$ is the sampled discrete-time channel response and $w[k]$ denotes discrete-time noise.

Frequency selectivity and dispersion, captured by the amplitude and phase of $H(f)$, manifest in the time domain as inter-symbol interference (ISI), thereby motivating receiver equalization to compensate for amplitude distortion and phase-induced temporal spreading. A detailed characterization of these effects is therefore essential to enable informed design choices in bandwidth selection, waveform shaping, and equalization strategy, ultimately supporting communication-system optimization. 

\section{Dispersive Electromagnetic Modeling of Biological Tissues}
\label{sec:tissue_properties}

The frequency-dependent electromagnetic response of biological tissues is modeled using the Cole--Cole formulation~\cite{gabriel1996dielectric}, which captures dispersive dielectric behavior over the kHz–GHz range and is widely adopted in wearable communication modeling~\cite{pun2011quasistatic,modak2022biophysical}. The complex relative permittivity is
\begin{equation}
\hat{\varepsilon}_r(f)=\varepsilon_{\infty}
+ \sum_{n=1}^{N}
\frac{\Delta \varepsilon_n}
     {1+\left(j2\pi f \tau_n\right)^{1-\beta_n}}
+ \frac{\sigma_i}{j2\pi f \varepsilon_0},
\label{eq:cole-cole}
\end{equation}
where $\varepsilon_{\infty}$ denotes the high-frequency permittivity,
$\Delta\varepsilon_n$ and $\tau_n$ represent the strength and characteristic time of the $n$-th relaxation process,
$\beta_n\in[0,1]$ models dispersion broadening,
$\sigma_i$ is the static ionic conductivity,
and $\varepsilon_0$ is the vacuum permittivity. This formulation incorporates the main polarization mechanisms occurring in biological media (e.g., Maxwell–Wagner interfacial effects and dipolar relaxations), making both effective permittivity and conductivity intrinsically frequency dependent.

In the electro-quasistatic GC band (10~kHz–1~MHz), current transport is governed by the complex conductivity
\begin{equation}
\hat{\sigma}(f)
=
\sigma_i
+
j2\pi f \varepsilon_0
\left(
\varepsilon_{\infty}
+
\sum_{n=1}^{N}
\frac{\Delta \varepsilon_n}
     {1+\left(j2\pi f \tau_n\right)^{1-\beta_n}}
\right),
\label{eq:sigma_complex}
\end{equation}
which directly enters the quasi-static field equation~\cite{pun2011quasistatic}
\begin{equation}\label{eq:qs_field}
\nabla \cdot \left[
\hat{\sigma}(f)\,\nabla \Phi(\mathbf{r},f)
\right]=0,
\end{equation}
governing the spatial electric potential $\Phi$ within the multilayer anatomical model and determines both the body input impedance $Z_{\mathrm{body,TX}}(f)$ and the transfer impedance $Z_{\mathrm{tr}}(f)$ introduced in the previous section.

Because $\hat{\sigma}(f)$ is complex and dispersive, each tissue layer exhibits both resistive and reactive behavior~\cite{gabriel1996dielectric}. In the lower portion of $\mathcal{B}$, conduction dominates and tissues behave predominantly resistively, consistent with electro-quasistatic intrabody models~\cite{pun2011quasistatic,modak2022biophysical}. As frequency increases, displacement currents associated with the permittivity term become progressively significant, introducing capacitive effects and frequency-dependent phase rotation~\cite{gabriel1996dielectric}. These dispersive properties ultimately shape the frequency dependence of $Z_{\mathrm{TX}}(f)$ and $Z_{\mathrm{tr}}(f)$, thereby determining attenuation, phase distortion, and the effective bandwidth of galvanic wearable communication links.

\section{Electrode Interface Modeling and Material Effects}
\label{sec:electrodes}

Electrode–tissue interface engineering strongly affects the electrical behavior of galvanic wearable links. Material choice, conductive gel, and mechanical stabilization determine injection impedance, boundary conditions, and long-term stability~\cite{wegmueller2010signal,callejon2013galvanic,modak2022biophysical}. Therefore, the electrode stack must be considered an integral part of the channel, not an ideal contact.

The differential electrode impedance $Z_e(f)$ in \eqref{eq:Ze_full} contributes to the input impedance $Z_{\mathrm{TX}}(f)$ in \eqref{eq:Zin} and affects the transfer impedance $Z_{\mathrm{tr}}(f)$ in \eqref{eq:Ztr_decomp}. Because $Z_{\mathrm{TX}}(f)$ sets current injection efficiency and $Z_{\mathrm{tr}}(f)$ governs tissue coupling, interface properties directly shape the channel response $H(f)$, influencing attenuation, phase, and broadband stability.

\subsection*{Electrode Materials}

We consider four materials commonly employed in wearable bioelectronic systems: Ag/AgCl, copper (Cu), platinum (Pt), and carbon-based electrodes. Their electrochemical behavior determines the effective charge-transfer resistance $R_c$ and double-layer capacitance $C_{dl}$~\cite{grimnes2014bioimpedance}.

Ag/AgCl electrodes exhibit quasi-reversible electrode kinetics due to the Ag/AgCl redox couple, resulting in low charge-transfer resistance and relatively stable interface potentials~\cite{wegmueller2010signal}. In contrast, Pt and Cu behave predominantly as polarizable electrodes in physiological environments, where current injection is governed mainly by double-layer charging with limited faradaic exchange. Carbon electrodes show intermediate behavior, with morphology-dependent capacitance and moderate polarization impedance.

Accordingly,
\begin{itemize}
\item Ag/AgCl typically exhibits lower $R_c$ and moderate $C_{dl}$;
\item Pt and Cu present higher effective polarization impedance (dominantly capacitive at low frequency);
\item Carbon electrodes exhibit material- and surface-dependent $C_{dl}$ and moderate $R_c$.
\end{itemize}

These differences directly modify $Z_e(f)$ and therefore alter injection efficiency and phase characteristics in the GC link.

\subsection*{Interface Configurations}

A rigorous characterization of GC links must explicitly account for the electrode–tissue interface, which governs current injection, boundary conditions, and field distribution. To systematically quantify its impact, four configurations are evaluated: (i) bare contact, (ii) foam only, (iii) gel only, and (iv) foam + gel.

Without gel, the skin stratum corneum acts as a high-resistivity barrier, increasing $R_c$ and inducing strongly capacitive behavior at low frequencies~\cite{grimnes2014bioimpedance}. This leads to large $|Z_e(f)|$ and pronounced phase rotation, limiting injection efficiency. Conductive gel introduces an admittance term $\sigma_g + j2\pi f\varepsilon_g$, which reduces the real part of $Z_e(f)$ and strengthens capacitive coupling. By enhancing ionic conduction across the stratum corneum and increasing the effective microscopic contact area~\cite{wegmueller2010signal,callejon2013galvanic}, gel lowers $|Z_e(f)|$ over the low-to-mid band $\mathcal{B}$, thereby improving current injection and mitigating phase distortion.

Foam primarily improves mechanical conformity and hydration stability. Although it does not substantially reduce intrinsic $R_c$, it stabilizes the contact geometry, decreasing temporal fluctuations in $R_c$ and $C_{dl}$. Foam also slightly alters parasitic capacitances $C_{\mathrm{par}}(f)$, leading to minor high-frequency variations.

Because $Z_e(f)$ defines the boundary conditions of the quasi-static field equation in \eqref{eq:qs_field}, interface engineering affects not only $Z_{\mathrm{TX}}(f)$ but also the spatial distribution of current within tissue. Variations in electrode configuration therefore modify dominant current paths $\Gamma$, alter the mutual impedance $Z_{\mathrm{tr}}(f)$, and ultimately reshape GC channel attenuation and phase characteristics.

\section{Physics-Based Galvanic-Coupling digital human twin}
\label{sec:digital_twin}
We develop a physics-based, parametric human DT that maps anatomical, material, geometrical, and interface parameters into a communication-consistent channel representation. The model is implemented in COMSOL Multiphysics\textsuperscript{\textregistered}~\cite{multiphysics1998introduction} and solved in the frequency domain under the quasi-static formulation introduced in Section~\ref{sec:tissue_properties}. 

Unlike prior GC modeling approaches that primarily report attenuation trends or impedance values~\cite{callejon2014finiteelement,modak2022biophysical}, the proposed DT directly computes the measurable transfer function $H(f)$ and extracts dispersion-aware communication metrics (path loss, phase, group delay) across the GC band. This enables systematic design-space exploration and waveform-level optimization.

\subsection{Parametric digital human twin Formulation}

Let $\Theta =
\left\{
\hat{\sigma}_i(f),\;
d_l,\;
d_r,\;
A,\;
\mathcal{I}
\right\}$ denote the set of physical parameters describing the wearable propagation environment, where $\hat{\sigma}_i(f)$ is the complex tissue conductivity (Cole–Cole based), $d_l$ and $d_r$ define electrode spacing, $A$ is the contact area, and $\mathcal{I}$ represents the electrode-interface stack (material, gel, foam configuration).

The digital human twin is formalized as the operator
\begin{equation}
\mathcal{T}:\Theta \rightarrow H_{\mathrm{DT}}(f),
\end{equation}
where $H_{\mathrm{DT}}(f)$ is the differential voltage transfer function defined in~\eqref{eq:channel}. From $H_{\mathrm{DT}}(f)=\alpha_{\mathrm{DT}}(f)e^{-j\phi_{\mathrm{DT}}(f)}$, communication-relevant metrics are derived directly. In particular, attenuation follows from $|H_{\mathrm{DT}}(f)|$ as in \eqref{eq:pathloss}, while the phase response $\phi_{\mathrm{DT}}(f)$ determines the phase delay
\begin{equation}\label{eq:phasedelay}
\tau_{p,\mathrm{DT}}(f) = -\frac{\phi_{\mathrm{DT}}(f)}{2\pi f},
\end{equation}
and the group delay
\begin{equation}\label{eq:groupdelay}
\tau_{g,\mathrm{DT}}(f) = -\frac{1}{2\pi}\frac{d\phi_{\mathrm{DT}}(f)}{df}.
\end{equation}
These quantities quantify frequency-dependent phase rotation and dispersion, and therefore provide a direct link between the physics-based model and waveform-level distortion analysis.

This formulation ensures that the DT produces quantities that are immediately usable for communication-system analysis, while remaining fully grounded in tissue electromagnetics.

\subsection{Electromagnetic Governing Model}

Within the GC band (10~kHz–1~MHz), the electromagnetic wavelength in tissue greatly exceeds device dimensions, justifying the electro-quasistatic approximation~\cite{pun2011quasistatic,maity2019biophysical}. The potential distribution is governed by the frequency-domain form of charge conservation in \eqref{eq:qs_field}, which is solved numerically over the multilayer anatomical domain using the \textit{Electric Currents} interface.

For each frequency point, the solver computes the scalar potential field $\Phi(\mathbf{r},f)$. The differential voltages at Tx and Rx electrode pairs are extracted directly, enabling computation of $H_{\mathrm{DT}}(f)$ without reliance on lumped approximations.
\begin{figure}[!t]  
    \centering  \includegraphics[width=\columnwidth]{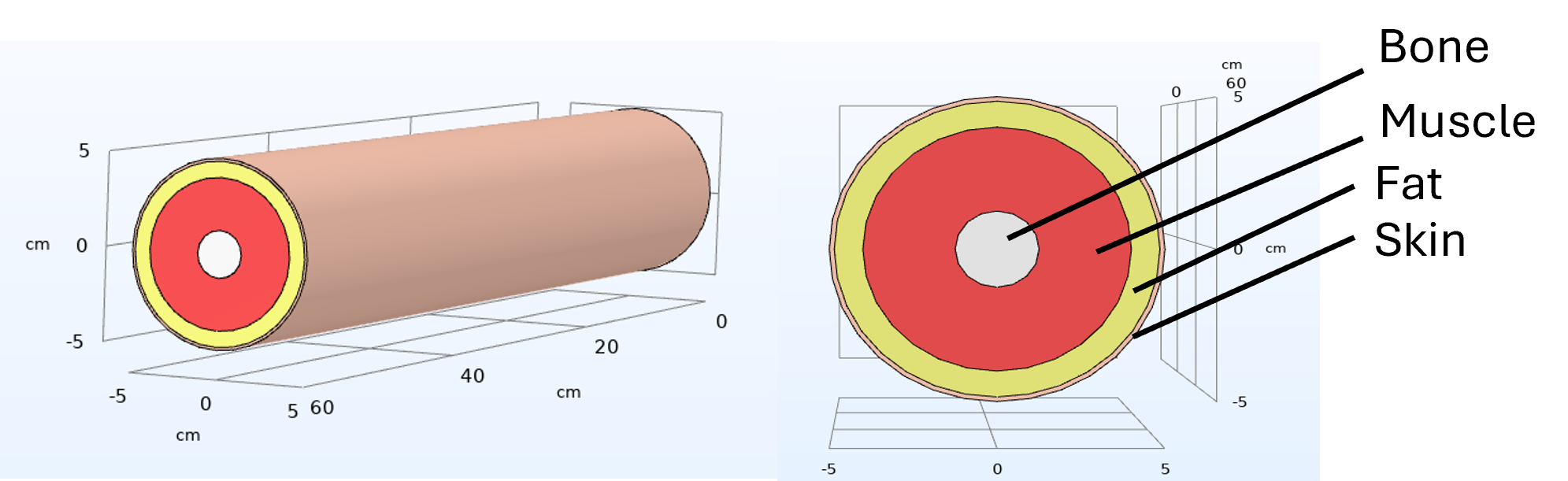}
    \caption{Concentric-layer arm model}
    \label{fig:arm}
\end{figure}
\subsection{Anatomical and Material Representation}

The arm is modeled as a concentric multilayer cylinder (skin, fat, muscle, bone) as in Fig. \ref{fig:arm}, consistent with established GC simulations~\cite{ahmed2019simulation}. Dispersive dielectric properties follow Cole–Cole parameters reported in Table~\ref{table:permittivity}~\cite{gabriel1996dielectric} and the arm parameters are detailed in Table~\ref{tab:arm_geometry}. 

The model is embedded in an air domain to ensure physically consistent boundary conditions and proper field decay. This prevents artificial reflections and preserves quasi-static field continuity.
\begin{table*}[!]
\centering
\caption{Cole-Cole parameters for selected biological tissues}
\label{table:permittivity}
\begin{tabular}{lcccccccccccccc}
\toprule
\textbf{Tissue type} & $\varepsilon_\infty$ & $\Delta\varepsilon_1$ & $\tau_1$ (ps) & $\beta_1$ & $\Delta\varepsilon_2$ & $\tau_2$ (ns) & $\beta_2$ & $\Delta\varepsilon_3$ & $\tau_3$ ($\mu$s) & $\beta_3$ & $\Delta\varepsilon_4$ & $\tau_4$ (ms) & $\beta_4$ & $\sigma$ \\
\midrule
Bone        & 2.5 & 10.0 & 13.26 & 0.20 & 180         & 79.58  & 0.20 & $5.0 \times 10^3$ & 159.15 & 0.20 & $1.0 \times 10^3$ & 15.915 & 0.00 & 0.0200 \\
Fat         & 2.5 & 3.0  & 7.96  & 0.20 & 15          & 15.92  & 0.10 & $3.3 \times 10^4$ & 159.15 & 0.05 & $1.0 \times 10^7$ & 7.958  & 0.01 & 0.0100 \\
Muscle      & 4.0 & 50.0 & 7.23  & 0.10 & 7000        & 353.68 & 0.10 & $1.2 \times 10^6$ & 318.31 & 0.10 & $2.5 \times 10^7$ & 2.274  & 0.00 & 0.2000 \\
Skin        & 4.0 & 32.0 & 7.23  & 0.00 & 1100        & 32.48  & 0.20 & 0.0               & 0.0    & 0.0  & 0.0               & 0.0    & 0.0  & 0.0002 \\
\bottomrule
\end{tabular}
\end{table*}
\begin{table}[!t]
\centering
\caption{Geometric parameters of the multilayered cylindrical arm model \cite{ahmed2019simulation}.}
\label{tab:arm_geometry}
\begin{tabular}{|l|c|l|}
\hline
\textbf{Model Component} & \textbf{Value }\\
\hline
Skin layer thickness      & 0.126 cm   \\
Fat layer thickness       & 0.580 cm  \\
Muscle layer thickness    & 1.550 cm \\
Bone layer thickness      & 1.244 cm    \\
Arm external radius    & 3.5 cm  \\
Arm length              & 60 cm  \\
Air domain radius         & 600 cm \\
\hline
\end{tabular}
\end{table}
\subsection{Electrode and Interface Modeling}

Surface-mounted circular electrodes are placed on the skin layer to emulate practical GC configurations. A constant contact area $A$ is maintained across simulations to isolate material and interface effects.

Electrode materials (Cu, Pt, Ag/AgCl, carbon) and interface stacks (bare, gel, foam, foam+gel) are explicitly modeled as volumetric layers rather than lumped impedances (Tab. \ref{tab:electrode_properties} and \ref{tab:electrode_geometry}). This ensures that the effective interface impedance $Z_e(f)$ emerges naturally from the field solution, allowing direct evaluation of how interface engineering modifies $Z_{\mathrm{eq}}(f)$.

\subsection{Excitation and Boundary Conditions}

A differential sinusoidal current source ($I_0=1~\mathrm{mA}$) excites the Tx dipole, consistent with experimental GC practice~\cite{wegmueller2009galvanic}. The Rx electrodes are modeled as floating terminals to approximate high-input-impedance sensing and avoid artificial loading. Impedance boundary conditions are applied at domain extremities to suppress non-physical reflections.

\begin{table}[!b]
\centering
\caption{Electrical properties of electrode and interface materials.}
\label{tab:electrode_properties}
\begin{tabular}{|l|c|c|}
\hline
\textbf{Material} & \textbf{Conductivity (S/m)} & \textbf{Relative Permittivity} \\
\hline
Copper (Cu)            & $5.8 \times 10^{7}$  & 1  \\
Platinum (Pt)          & $9.4 \times 10^{6}$  & 1  \\
Ag/AgCl                & $1.5 \times 10^{4}$  & 12 \\
Carbon                 & $1.0 \times 10^{4}$  & 10 \\
Conductive hydrogel    & 0.5                  & 78 \\
Foam backing           & $1 \times 10^{-15}$  & 1.5 \\
\hline
\end{tabular}
\end{table}

\begin{table}[!b]
\centering
\caption{Electrode geometry parameters}
\label{tab:electrode_geometry}
\setlength{\tabcolsep}{4pt}
\begin{tabular}{lcc}
\toprule
\textbf{Component} & \textbf{Radius [cm]} & \textbf{Thickness [mm]} \\
\midrule
Conductor & $R_c=1$ & 2.0 \\
Foam      & $R_f=R_c+0.4$ & 3.7 \\
Gel       & $R_g=R_c+0.2$ & 0.2 \\
\midrule
$d_r$ & 4 & -- \\
$d_l$ & 10 (long.), 7 (rad.) & -- \\
\bottomrule
\end{tabular}
\end{table}
\subsection{Digital Human Twin Capabilities}

By directly translating dispersive tissue electromagnetics into communication-compatible transfer functions, the proposed DT establishes a rigorous bridge between physics-based modeling and system-level design.

Prior GC studies have provided valuable electromagnetic and circuit-level analyses~\cite{callejon2012distributed,callejon2014finiteelement,pun2011quasistatic,modak2022biophysical}, primarily focusing on attenuation trends, impedance extraction, or loading effects. However, these approaches largely rely on simulation or measurement for descriptive validation, without yielding a dispersion-aware, design-oriented channel representation suitable for communication analysis.

In contrast, the proposed DT computes the full complex transfer function $H(f)$, capturing amplitude, phase, and group delay while volumetrically resolving electrode–tissue interface physics. The framework systematically incorporates both interface configurations and physiologically realistic Tx–Rx separations along the arm (e.g., wrist–forearm, wrist–upper arm, forearm–upper arm), covering longitudinal and radial placements consistent with anatomical constraints.

As a result, the parametric channel formulation enables controlled exploration of anatomical and interface variability and produces physics-consistent channel models directly usable for waveform design, bandwidth optimization, and system-level performance evaluation.


\begin{table}[!t]
\centering
\caption{Main In-Vivo Experimental Parameters (Galvanic Coupling)}
\label{tab:invivo_parameters}
\begin{tabular}{ll}
\hline
\textbf{Parameter} & \textbf{Value} \\
\hline
PN sequence degree & $m = 14$ \\
Chip duration $T_c$ & $5.2~\mu$s \\
Effective bandwidth & $\approx 96$~kHz \\
Sampling frequency $f_s$ & $192$~kHz \\
ADC resolution & 16 bits \\
Transmit power & $\approx 10~\mu$W \\
Measured SNR & $\approx 20$~dB \\
Electrode type & Ag/AgCl surface electrodes \\
Electrode radius & 1~cm \\
Inter-electrode spacing & 4~cm \\
Tx--Rx separation $d_{\mathrm{tx-rx}}$ & 10 ( Longitudinal), 7 (Radial)~cm \\
Repetitions per setup & $>40$ \\
\hline
\end{tabular}
\end{table}

\section{Experimental Validation and FEM--Measurement Alignment}
\label{sec:validation}

The proposed DT is validated through controlled in-vivo GC measurements performed on the human forearm. The experimental methodology follows an impulse-response-based channel sounding approach consistent with recent GC characterization studies~\cite{vizziello2023experimental}, ensuring methodological alignment between simulation and measurement.

The experimental measurements were conducted on healthy adult volunteers using non-invasive wearable electrodes. According to institutional guidelines, this type of non-invasive experimental activity does not require formal approval by an Institutional Review Board. All participants were informed about the nature and purpose of the study and provided written informed consent prior to participation. All procedures complied with relevant safety standards.

\subsection{Measurement Campaign and Instrumentation}

The experimental campaign targets electro-quasistatic GC operation below 100 kHz. Channel reconstruction employs correlation-based sounding using a maximal-length pseudonoise (PN) sequence ($m=14$) transmitted in baseband with chip duration $T_c=5.2,\mu s$, corresponding to an excitation bandwidth of $\approx96$ kHz. Signals are acquired at $f_s=192$ kHz with 16-bit resolution. The channel impulse response (CIR) is obtained by correlating the received waveform with the known PN sequence, and the frequency response follows via Fourier transform.

To suppress unintended common-ground return paths, the transmitter (battery-powered) and receiver (separate workstation) are electrically isolated. Differential injection and acquisition are performed using commercial audio interfaces. The transmit power is on the order of $10\,\mu\mathrm{W}$, ensuring negligible tissue heating and compliance with low-power wearable constraints, with a measured SNR of approximately $20\,\mathrm{dB}$.

Surface-mounted Ag/AgCl electrodes (1\,cm radius, 4\,cm intra-pair spacing) with gel and foam backing provide stable boundary conditions. Longitudinal and radial Tx–Rx placements are evaluated using physiologically realistic separations: $l_d=10\,\mathrm{cm}$ (longitudinal) and $l_d=7\,\mathrm{cm}$ (radial), representative of practical forearm deployments. The radial spacing approximates the typical forearm diameter, ensuring realistic cross-sectional current confinement, while the larger longitudinal spacing maintains spatially distinct injection and sensing regions, reducing near-field coupling and promoting distributed volumetric conduction. For each configuration, more than 40 acquisitions are ensemble-averaged to quantify statistical variability (Tab.~\ref{tab:invivo_parameters}).

\begin{figure}[!t]  
    \centering  \includegraphics[width=\columnwidth]{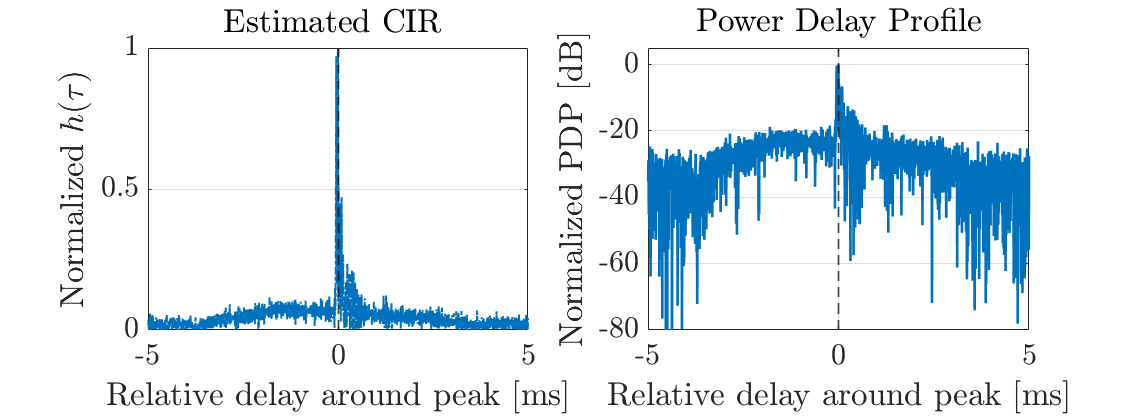}
    \caption{Estimated CIR and normalized power delay profile (PDP) for longitudinal configuration.}
    \label{fig:CIR}
\end{figure}

The reconstructed CIRs in Fig.~\ref{fig:CIR} show a single dominant component with no resolvable secondary peaks within the considered bandwidth. The normalized PDPs confirm that the received energy is concentrated around the main delay, with no distinct delayed contributions. The slight temporal spreading around the peak is attributable to finite bandwidth and tissue dispersion, rather than multipath propagation. This behavior is consistent with electro–quasistatic volumetric conduction and agrees with prior measurements~\cite{vizziello2023experimental}. Similar results are obtained for both longitudinal and radial configurations (the latter omitted for brevity).

\subsection{Calibration and FEM--Measurement Mapping}
\label{sec:dt_validation}

To ensure a fair and physically consistent comparison between simulations and measurements, a structured calibration and alignment procedure is adopted. The objective is to isolate intrinsic channel behavior from instrumentation-induced distortions.

First, the complex transfer function is reconstructed as
\[
\hat{H}(f)=\frac{S_{yx}(f)}{S_{xx}(f)},
\]
where $S_{yx}(f)$ and $S_{xx}(f)$ denote the measured cross-spectrum and the PN auto-spectrum, respectively. All spectra are interpolated onto a common frequency grid and restricted to the quasi-static band considered in the FEM model.

\subsubsection*{Instrumentation Offset Compensation}

Acquisition chains introduce systematic, frequency-independent distortions arising from front-end electronics and cabling, primarily manifesting as constant gain scaling and residual phase offsets. To compensate for these effects, a multiplicative magnitude correction factor is estimated within a central calibration band characterized by high signal-to-noise ratio and minimal spectral structure. The factor is computed using a robust median estimator to mitigate the influence of outliers and narrowband fluctuations, and is applied uniformly across all repetitions, assuming instrumentation stability over the acquisition set.

After phase unwrapping, a constant (frequency-independent) phase offset is estimated over the same calibration band and subtracted from the entire spectrum. This correction removes deterministic interface-induced phase bias without altering intrinsic frequency-dependent phase trends.

By restricting the compensation to frequency-independent components only, the procedure preserves the genuine spectral amplitude and phase behavior of the measured system while eliminating systematic acquisition artifacts.

\subsubsection*{Parameter-Level Digital Human Twin Alignment}

After instrumentation compensation, alignment between measurements and the DT is conducted at the parameter level rather than by directly fitting the transfer function. Anatomical and interface parameters in $\Theta$ are set to match the experimental configuration, and DT response is then evaluated without further tuning, ensuring that any observed agreement arises from physically grounded parameterization.

\subsubsection*{Dispersion Metric Extraction}

The measured phase is unwrapped and regularized using a Savitzky–Golay filter to suppress high-frequency noise while preserving the global trend. Phase delay $\tau_P$ and group delay $\tau_G$ are then computed according to the definitions in \eqref{eq:phasedelay} and \eqref{eq:groupdelay}. Because group delay involves numerical differentiation, it is inherently more sensitive to residual noise; therefore, validation emphasis is placed on attenuation and phase delay.

\subsubsection*{Statistical Similarity Assessment}

For each frequency point, replicate-wise mean and standard deviation are computed. The consistency between the DT and the measured channel responses is evaluated using complementary statistical indicators: the root-mean-square error (RMSE), the mean absolute error (MAE), the bias (mean signed deviation), the Pearson correlation coefficient (Corr), defined with respect to the dynamic range of the DT reference curve. Correlation for attenuation is computed after mild moving-average smoothing to emphasize trend similarity rather than stochastic ripple, which is largely driven by contact variability in low-frequency GC measurements.

\begin{figure}[!t]
\centering
\begin{subfigure}[t]{0.80\columnwidth}
    \centering
    \includegraphics[width=\linewidth]{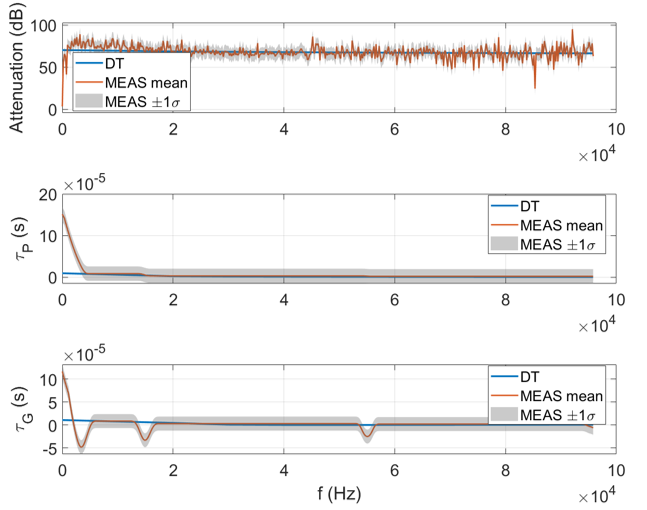}
    \caption{Longitudinal configuration}
    \label{fig:val_long}
\end{subfigure}

\vspace{0.2em}

\begin{subfigure}[t]{0.80\columnwidth}
    \centering
    \includegraphics[width=\linewidth]{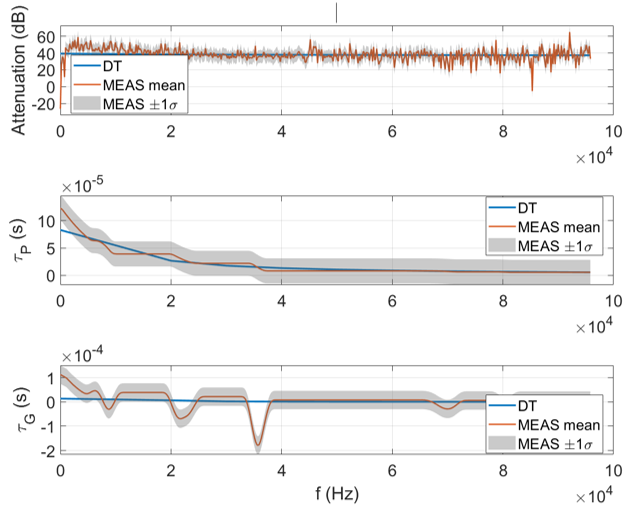}
    \caption{Radial configuration}
    \label{fig:val_rad}
\end{subfigure}

\caption{Digital human twin versus in-vivo measurements for galvanic coupling under (a) longitudinal and (b) radial electrode configurations. For each configuration, attenuation, phase delay $\tau_P$, and group delay $\tau_G$ are shown as mean and $\pm1\sigma$ across repeated acquisitions over the quasi-static band (0--100~kHz).}
\label{fig:validation_overall}
\end{figure}

\begin{table}[!t]
\centering
\caption{digital human twin vs.\ Measurements: similarity metrics for longitudinal and radial configurations (in-vivo, galvanic coupling).}
\label{tab:similarity_overall}
\resizebox{\columnwidth}{!}{%
\begin{tabular}{llrrr}
\hline
\textbf{Configuration} & \textbf{Quantity} & \textbf{RMSE} & \textbf{MAE} & \textbf{Corr} \\
\hline
\multirow{3}{*}{Longitudinal} 
& Attenuation (dB) & 5.86 & 4.34 & 0.74 \\
& Phase delay (s)  & $1.82\times 10^{-6}$ & $1.66\times 10^{-6}$ & 0.85 \\
& Group delay (s)  & $8.28\times 10^{-6}$ & $4.31\times 10^{-6}$ & -- \\
\hline
\multirow{3}{*}{Radial} 
& Attenuation (dB) & 6.00 & 4.44 & 0.75 \\
& Phase delay (s)  & $5.11\times 10^{-6}$ & $4.08\times 10^{-6}$ & 0.91 \\
& Group delay (s)  & $4.08\times 10^{-5}$ & $2.50\times 10^{-5}$ & -- \\
\hline
\end{tabular}}
\end{table}
Figures~\ref{fig:validation_overall} and Table~\ref{tab:similarity_overall} compare the DT with in-vivo galvanic measurements for both longitudinal and radial configurations. The attenuation mismatch remains within $\approx6$ dB RMSE in both cases, consistent with expected variability from electrode–skin impedance, slight placement deviations, and subject-specific tissue properties in low-frequency wearable links. To mitigate stochastic magnitude fluctuations, attenuation correlation is computed on window-averaged responses, emphasizing trend consistency rather than pointwise ripple. This yields correlation coefficients of $\approx0.74–0.75$, confirming the visual agreement of the attenuation profiles.

Phase-delay agreement shows microsecond-level absolute errors and high Pearson correlations (0.85–0.91), indicating that the DT accurately reproduces the dominant frequency-dependent phase evolution in the quasi-static band. Group delay is reported for completeness but not used as a primary validation metric, since phase differentiation inherently amplifies noise and smoothing artifacts.

Overall, attenuation and phase-delay metrics provide consistent, physically meaningful evidence that the proposed DT captures both amplitude behavior and dispersive characteristics of quasi-static wearable galvanic channels across propagation directions and electrode configurations.

\section{Numerical Results and Discussion}\label{sec:num_results}
This section systematically characterizes the wearable GC channel under different propagation geometries and signaling regimes. Numerical results are obtained using $d_r=4$ cm (intra-pair spacing) and physiologically realistic Tx–Rx separations of $d_l=10$ cm (longitudinal) and $d_l=7$ cm (radial), with electrode and foam geometry matched to the measurement setup. Attenuation, phase delay, and group delay are analyzed and normalized with respect to $d_l$, enabling direct comparison between longitudinal and radial configurations. Their contrast isolates the impact of anatomical anisotropy, including muscle fiber orientation and bone proximity, on conductive transport.

Two complementary regimes are considered. In the narrowband case, where the signal bandwidth lies within the channel coherence bandwidth, the channel is treated as frequency-flat and evaluated versus carrier frequency, enabling controlled assessment under quasi-static conditions. In the wideband regime, the signal spans frequencies over which the response varies appreciably; frequency selectivity and dispersion therefore emerge, and joint amplitude–phase analysis reveals bandwidth-dependent distortion and delay variability.

By combining directional (longitudinal vs.\ radial) and bandwidth-dependent analyses, this section provides a physics-consistent interpretation of wearable GC behavior and establishes quantitative guidelines for bandwidth selection and transceiver design.
\subsection{Narrowband Condition}

\begin{table}[!t]
\centering
\caption{Normalized channel parameters for wearable galvanic coupling. 
$PL/d$ is the path loss per unit distance [dB/cm]; 
$k_p=\tau_P/d$ and $k_g=\tau_G/d$ are the phase- and group-delay coefficients [ns/cm].}
\label{tab:norm_channel_compact}
\scriptsize
\renewcommand{\arraystretch}{1.1}
\begin{tabular}{lcccccc}
\toprule
 & \multicolumn{2}{c}{$PL/d$ (dB/cm)} 
 & \multicolumn{2}{c}{$k_p$ (ns/cm)} 
 & \multicolumn{2}{c}{$k_g$ (ns/cm)} \\
\cmidrule(r){2-3} \cmidrule(r){4-5} \cmidrule(r){6-7}
Config & 10 kHz & 1 MHz & 10 kHz & 1 MHz & 10 kHz & 1 MHz \\
\midrule
Long–Cu   & 6.77 & 6.42 & 135 & 3 & 90  & -2 \\
Long–Pt   & 6.76 & 6.44 & 118 & 4 & 95  & -3 \\
Long–C    & 6.75 & 6.43 & 120 & 3 & 75  & -4 \\
Long–AgCl & 6.75 & 6.41 & 108 & 2 & 68  & -5 \\
\midrule
Rad–Cu    & 5.44 & 5.31 & 110 & 2 & 300 & -5 \\
Rad–Pt    & 5.42 & 5.30 & 205 & 3 & 265 & -6 \\
Rad–C     & 5.42 & 5.31 & 195 & 3 & 240 & -7 \\
Rad–AgCl  & 5.43 & 5.30 & 180 & 2 & 220 & -6 \\
\bottomrule
\end{tabular}
\end{table}

Table~\ref{tab:norm_channel_compact} reports normalized attenuation and delay coefficients for longitudinal and radial GC configurations under gel+foam conditions. The narrowband analysis isolates propagation-geometry effects under stabilized electrode boundaries.

The normalized path loss, defined as $PL/d$, is higher in the longitudinal case ($\approx\,$6.7 dB/cm) than in the radial case ($\approx\,$5.3 dB/cm), reflecting geometry-dependent current distributions and muscle anisotropy. Although normalization mitigates first-order distance effects, residual separation-dependent factors (e.g., effective conductive cross-section and boundary interaction) may persist. Thus, the lower radial $PL/d$ indicates geometry sensitivity rather than a placement recommendation. Across electrode materials, variations are marginal, confirming that under gel+foam stabilization the narrowband magnitude response is dominated by bulk tissue conduction.

The phase-delay coefficient $k_p=\tau_P/d_l$ decreases from 10 kHz to 1 MHz in both geometries, consistent with the transition from polarization-dominated conduction to increased displacement-current contribution. Its weak material dependence further supports bulk-admittance-driven phase evolution. The group-delay coefficient $k_g=\tau_G/d_l$ is positive at low frequency and approaches small negative values near 1 MHz, reflecting mild high-frequency phase flattening typical of dispersive electro-quasistatic media rather than anomalous propagation.

Overall, attenuation and delay metrics indicate a quasi-static, weakly dispersive channel over 10 kHz–1 MHz. For bandwidths within the coherence bandwidth, a frequency-flat model is adequate. From a design perspective, gel+foam conditioning renders electrode material a second-order factor in narrowband links, while carrier selection above a few hundred kHz reduces phase curvature and simplifies receiver equalization.
\begin{figure}[!t]
\centering
\begin{subfigure}[t]{0.9\columnwidth}
    \centering
    \includegraphics[width=\linewidth]{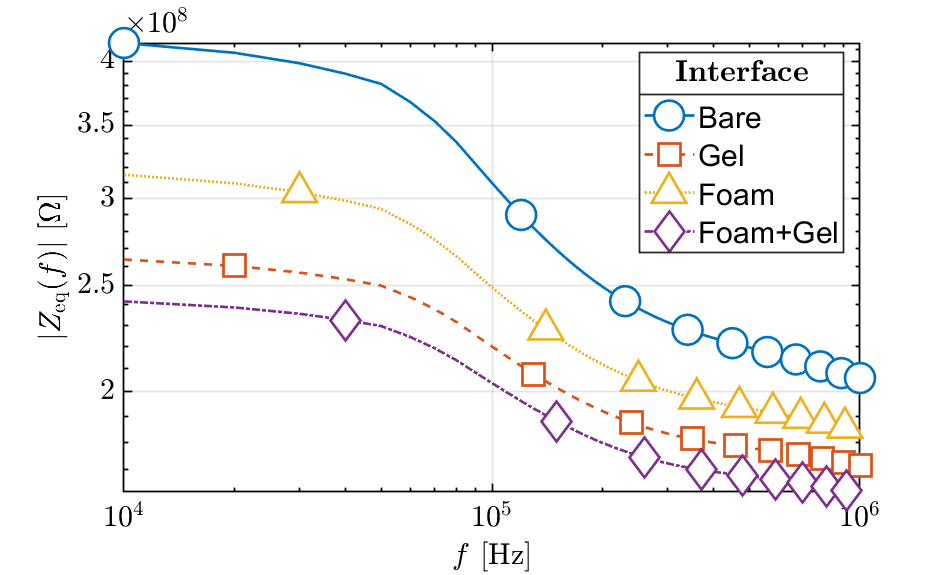}
    \caption{}
    \label{fig:Z_long}
\end{subfigure}

\vspace{0.2em}

\begin{subfigure}[t]{0.9\columnwidth}
    \centering
    \includegraphics[width=\linewidth]{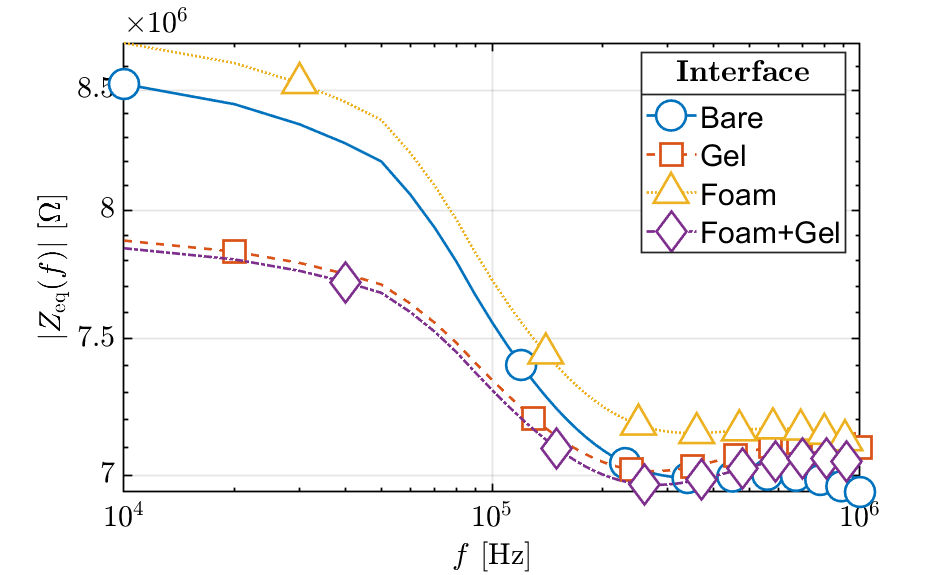}
    \caption{}
    \label{fig:Z_rad}
\end{subfigure}
\caption{Equivalent impedance $Z_{eq}(f)$ for (a) longitudinal and (b) radial configuration vs frequency for different AgCl electrodes Configurations}
\label{fig:Z_config}
\end{figure}

\begin{figure}[!t]
\centering
\begin{subfigure}[t]{0.90\columnwidth}
    \centering
    \includegraphics[width=\linewidth]{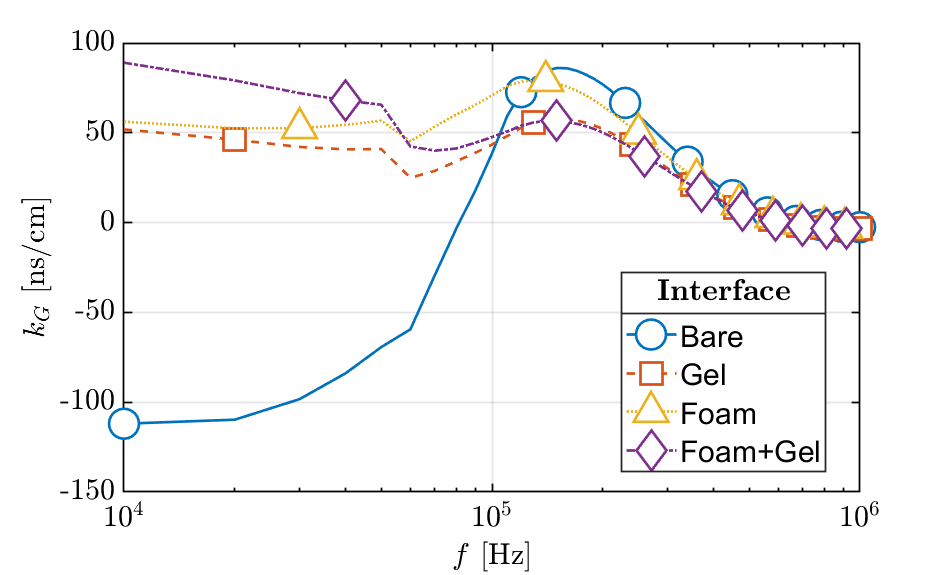}
    \caption{}
    \label{fig:kg_long}
\end{subfigure}

\vspace{0.2em}

\begin{subfigure}[t]{0.90\columnwidth}
    \centering
    \includegraphics[width=\linewidth]{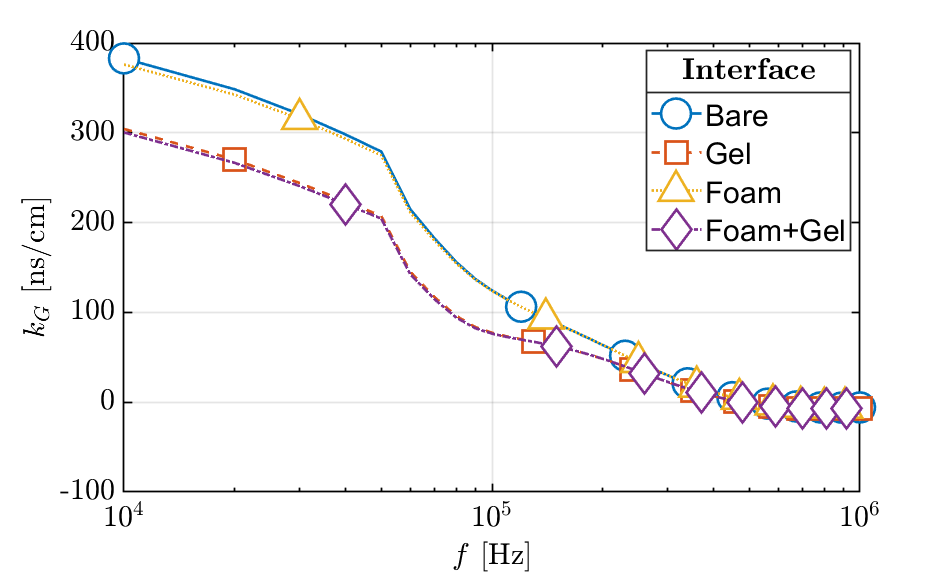}
    \caption{}
    \label{fig:kg_rad}
\end{subfigure}
\caption{Normalized group-delay coefficient $k_g$ as a function of frequency for (a) longitudinal and (b) radial galvanic configurations using Ag/AgCl electrodes under different interface conditions (bare, foam, gel, and foam+gel).}
\label{fig:kg_config}
\end{figure}

Figures~\ref{fig:Z_config} and \ref{fig:kg_config} report results obtained with Ag/AgCl surface electrodes to isolate interface-engineering effects under clinically relevant conditions. The equivalent impedance $Z_{\mathrm{eq}}(f)$ in \eqref{eq:pathloss} captures both bulk tissue conduction and electrode–skin boundary contributions. Attenuation is referenced to $Z_{\mathrm{ref}}=Z_L=10^5\,\Omega$, representative of a high-input-impedance wearable front-end.

Gel and foam conditioning substantially reduce $|Z_{\mathrm{eq}}(f)|$ at low frequencies, where polarization and contact impedance dominate. At higher frequencies, responses converge, indicating that volumetric tissue admittance progressively governs the channel and boundary effects diminish.
Propagation geometry induces systematic differences. The longitudinal case shows higher normalized impedance and stronger low-frequency dispersion, consistent with anisotropic muscle conductivity and fiber-aligned current flow. The radial configuration exhibits lower normalized attenuation, but is evaluated at a shorter $d_l$ (7 cm vs.\ 10 cm longitudinal), consistent with typical forearm diameters. Although normalization by $d_l$ mitigates first-order distance effects, residual separation-dependent factors and the presence of low-conductivity bone in radial paths, reducing effective conductive cross-section, also influence current distribution. Thus, the observed differences indicate geometry and anatomical sensitivity rather than a prescriptive placement rule.

The normalized group-delay coefficient $k_g=\tau_G/d$ in Fig.\ref{fig:kg_config} further characterizes dispersion. Large positive values at low frequency reflect phase curvature driven by polarization and capacitive boundary effects. With increasing frequency, $k_g$ decreases toward small negative values, indicating phase flattening in the electro-quasistatic regime as displacement-current contributions compensate low-frequency curvature. This transition reflects weak residual dispersion, not anomalous propagation.

Overall, the wearable GC channel remains quasi-static and weakly dispersive over 10 kHz–1 MHz. Interface conditioning stabilizes low-frequency behavior, while operation above a few hundred kHz reduces boundary sensitivity and supports simplified narrowband modeling with limited equalization requirements.

\subsection{Wideband Condition}

In the wideband regime, the signal bandwidth spans frequencies over which the transfer function $H(f)$ varies appreciably in both magnitude and phase. Frequency selectivity and dispersion therefore become non-negligible, and joint amplitude–phase distortion determines waveform integrity.

\begin{figure}[!t]  
    \centering  
    \includegraphics[width=0.9\columnwidth]{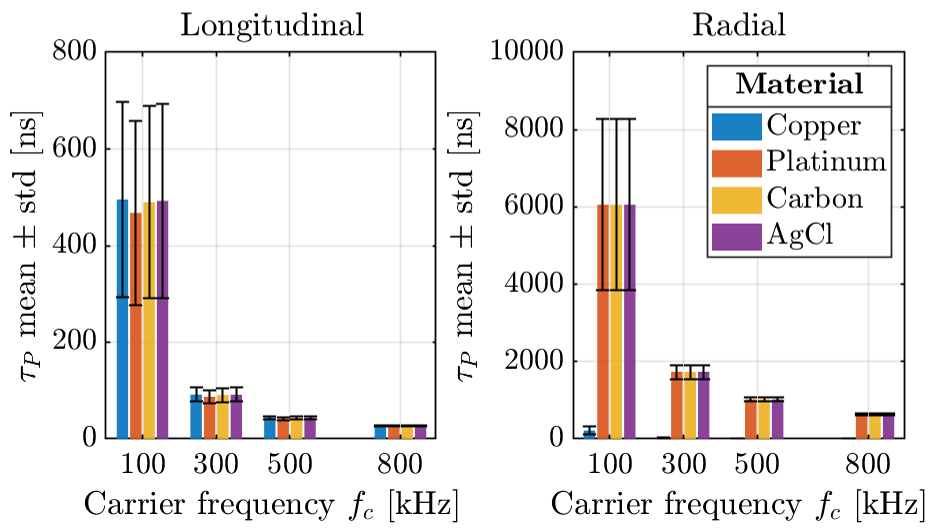}
    \caption{Mean phase delay $\tau_P$ (mean $\pm$ standard deviation) versus carrier frequency for longitudinal (left) and radial (right) configurations using a gel+foam interface with a fixed 100\,kHz bandwidth.}
    \label{fig:tauP_gelfoam}
\end{figure}

Figure~\ref{fig:tauP_gelfoam} reports the mean phase delay $\tau_P$ under gel+foam stabilization for varying carrier frequency with fixed bandwidth $\mathcal{B}=100$ kHz. In the longitudinal configuration, $\tau_P$ decreases monotonically with carrier frequency, indicating the transition from polarization-dominated conduction to increased displacement-current contribution. The radial case exhibits larger low-frequency delays, consistent with geometry-dependent current redistribution and greater sensitivity to anatomical heterogeneity. At higher frequencies, differences across materials and geometries diminish, confirming that bulk tissue admittance governs phase evolution as polarization effects weaken.

Wideband dispersion is quantified through intra-band variability metrics at fixed carrier frequency $f_c=500$ kHz, while the signal bandwidth is progressively increased.

\begin{figure}[!t]
\centering

\begin{subfigure}[t]{0.95\columnwidth}
    \centering
    \includegraphics[width=\linewidth]{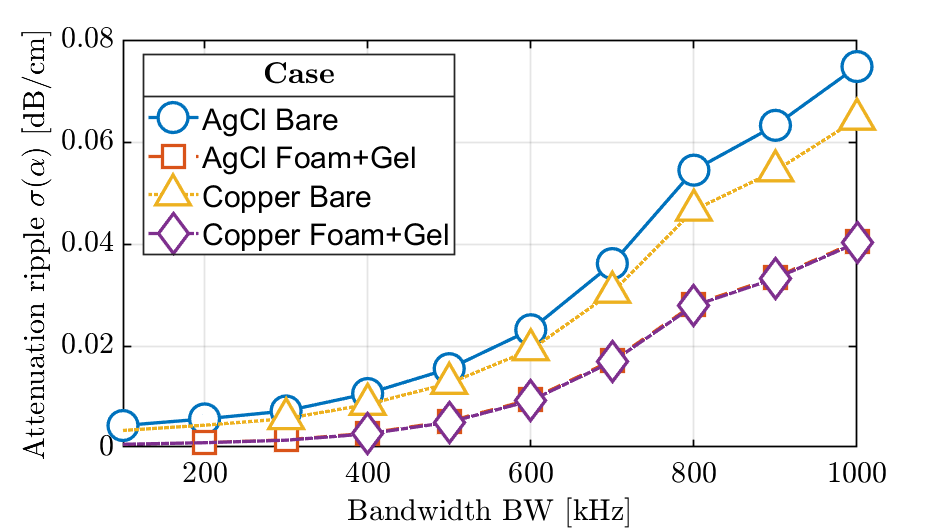}
     \caption{}
\end{subfigure}

\vspace{0.4em}

\begin{subfigure}[t]{0.95\columnwidth}
    \centering
    \includegraphics[width=\linewidth]{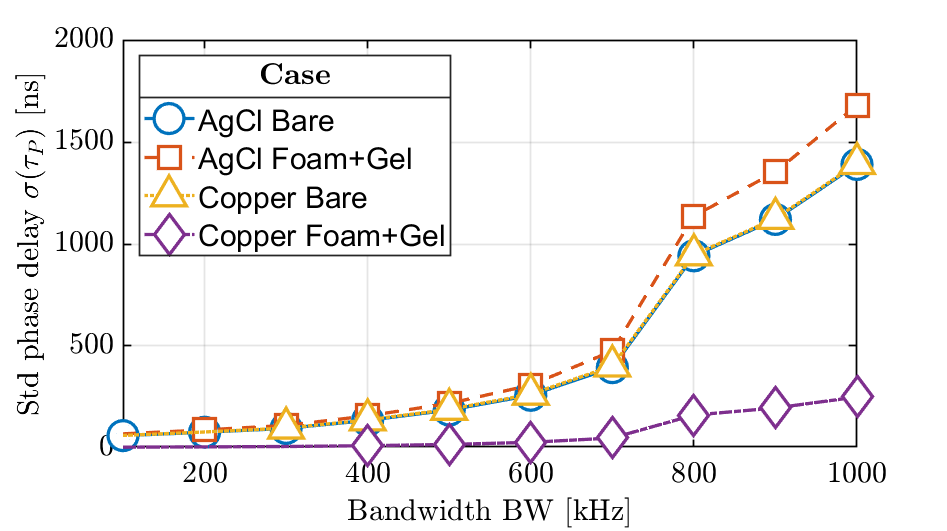}
    \caption{}
\end{subfigure}

\vspace{0.4em}

\begin{subfigure}[t]{0.95\columnwidth}
    \centering
    \includegraphics[width=\linewidth]{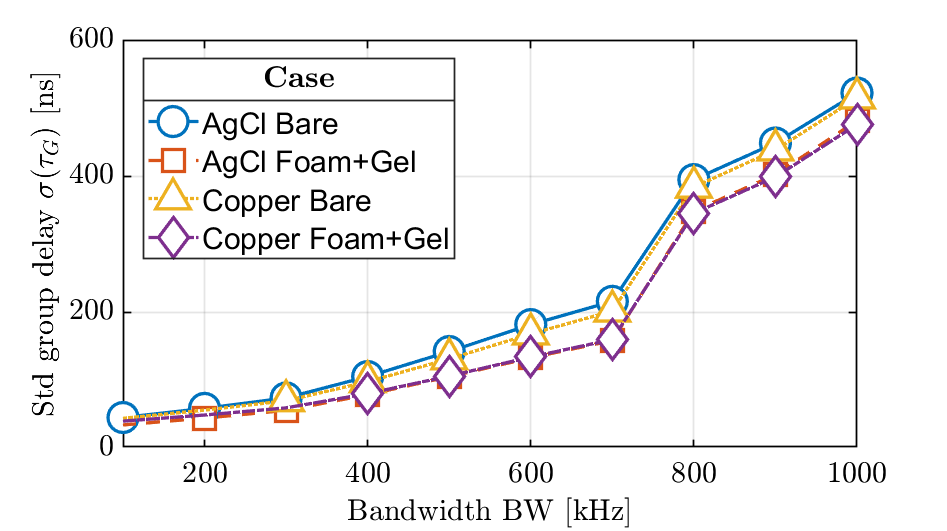}
    \caption{}
\end{subfigure}

\caption{Wideband dispersion metrics versus signal bandwidth at carrier frequency $f_c=500$ kHz for Ag/AgCl and copper electrodes under bare and foam+gel conditions: (a) attenuation ripple $\sigma(\alpha)$, (b) phase-delay spread $\sigma(\tau_P)$, and (c) group-delay spread $\sigma(\tau_G)$.}
\label{fig:bandwidth_dispersion}
\end{figure}

Figure~\ref{fig:bandwidth_dispersion}~(a) shows the standard deviation of attenuation ripple $\sigma(\alpha)$, defined relative to the mean attenuation over the band. As bandwidth expands, $\sigma(\alpha)$ increases systematically, indicating accumulation of frequency-dependent impedance variations into measurable amplitude distortion. Foam+gel conditioning consistently reduces ripple, improving spectral smoothness.

Figure~\ref{fig:bandwidth_dispersion}~(b) reports the standard deviation of phase delay $\sigma(\tau_P)$, which grows nonlinearly with bandwidth due to cumulative phase curvature. Bare electrodes exhibit markedly higher variability, especially above 600–700 kHz, whereas gel+foam stabilization mitigates timing fluctuations.

Figure~\ref{fig:bandwidth_dispersion}~(c) presents the standard deviation of group delay $\sigma(\tau_G)$, capturing higher-order phase dispersion. Its monotonic increase with bandwidth confirms enhanced sensitivity to phase nonlinearity in wideband operation. Across all metrics, interface conditioning reduces variability, highlighting the importance of boundary stabilization for temporal and spectral consistency.

Collectively, the results indicate that wideband GC performance is primarily limited by dispersion-induced amplitude and timing variability, rather than sharp frequency selectivity.

Three design implications emerge: (i) increasing bandwidth amplifies attenuation ripple and delay spread, tightening equalization and synchronization constraints; (ii) interface stabilization (gel+foam) significantly improves robustness by reducing spectral and temporal fluctuations; and (iii) operation in the mid-frequency range, above a few hundred kHz yet below pronounced attenuation growth, provides a favorable trade-off between power efficiency and dispersion control.

Therefore, wideband GC design requires joint optimization of bandwidth, interface conditioning, and transceiver complexity to preserve waveform integrity under electro-quasistatic conduction.

\section{Conclusions}\label{sec:conclusion}
This work presented a systematic characterization of wearable GC channels under both narrowband and wideband operation, supported by a calibrated physics-based digital human twin. Validated through in-vivo measurements, the DT enabled quantitative assessment of how propagation geometry, electrode–skin interface conditioning, and bandwidth jointly influence attenuation and dispersion, while serving as a predictive tool beyond purely empirical analysis. 
The results confirm electro–quasistatic, weakly dispersive behavior over 10~kHz--1~MHz. Attenuation is predominantly governed by geometry, whereas dispersion-induced variability increases with bandwidth. Interface stabilization (gel+foam) markedly reduces spectral ripple and delay variability, and operation above a few hundred kHz mitigates phase curvature, thereby supporting simplified modeling assumptions.
In conclusion, the DT-based framework establishes a physics-consistent and scalable foundation for model-driven design of wearable GC systems, enabling quantitative optimization across carrier frequency, bandwidth, power efficiency, and transceiver complexity.
\bibliographystyle{IEEEtran}
\bibliography{bibliography}
\end{document}